\def\DocType{sigconf} %
\documentclass[\DocType]{acmart}

\AtBeginDocument{%
  \providecommand\BibTeX{{%
    \normalfont B\kern-0.5em{\scshape i\kern-0.25em b}\kern-0.8em\TeX}}}

\copyrightyear{2022} 
\acmYear{2022} 
\setcopyright{rightsretained} 
\acmConference[CHI '22]{CHI Conference on Human Factors in Computing Systems}{April 29-May 5, 2022}{New Orleans, LA, USA}
\acmBooktitle{CHI Conference on Human Factors in Computing Systems (CHI '22), April 29-May 5, 2022, New Orleans, LA, USA}
\acmDOI{10.1145/3491102.3517682}
\acmISBN{978-1-4503-9157-3/22/04}

\usepackage{xspace}
\usepackage{calc}
\newcommand{\system}{ShapeFindAR\xspace}

\newcommand{\mmf}{MyMiniFactory\xspace}
\newcommand{\tv}{Thingiverse\xspace}

\begin{document}

\title[\system: Exploring In-Situ Spatial Search for Physical Artifact Retrieval using MR]{\system: Exploring In-Situ Spatial Search for Physical Artifact Retrieval using Mixed Reality}

\author{Evgeny Stemasov}
\email{evgeny.stemasov@uni-ulm.de}
\orcid{0000-0002-3748-6441}
\author{Tobias Wagner}
\email{tobias.wagner@uni-ulm.de}
\affiliation{%
  \institution{Institute of Media Informatics,\\Ulm University}
  \city{Ulm}
  \country{Germany}}

\author{Jan Gugenheimer}
\orcid{0000-0002-6466-3845}
\affiliation{%
  \institution{Télécom Paris -- LTCI, Institut Polytechnique de Paris}
  \city{Paris}
  \country{France}}
\email{jan.gugenheimer@telecom-paris.fr}

\author{Enrico Rukzio}
\email{enrico.rukzio@uni-ulm.de}
\affiliation{%
  \institution{Institute of Media Informatics,\\Ulm University}
  \city{Ulm}
  \country{Germany}}

\renewcommand{\shortauthors}{Stemasov, et al.}

\begin{abstract}
 Personal fabrication is made more accessible through repositories like Thingiverse, as they replace modeling with retrieval.
However, they require users to translate spatial requirements to keywords, which paints an incomplete picture of physical artifacts: proportions or morphology are non-trivially encoded through text only.
We explore a vision of in-situ spatial search for (future) physical artifacts, and present \system, a mixed-reality tool to search for 3D models using in-situ sketches blended with textual queries.
With \system, users search for geometry, and not necessarily precise labels, while coupling the search process to the physical environment (e.g., by sketching in-situ, extracting search terms from objects present, or tracing them).
We developed \system for HoloLens 2, connected to a database of 3D-printable artifacts.
We specify in-situ spatial search, describe its advantages, and present walkthroughs using \system, which highlight novel ways for users to articulate their wishes, without requiring complex modeling tools or profound domain knowledge.

\end{abstract}



\begin{teaserfigure}
    \centering
    \includegraphics[width=1\linewidth]{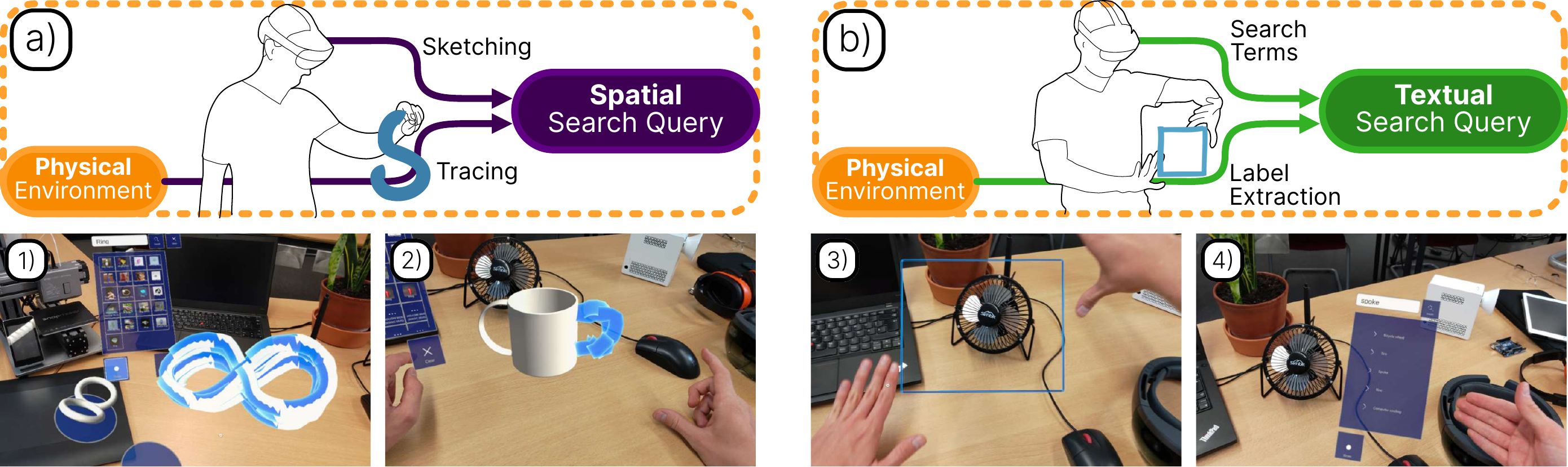}
    \Description{Figure 1.a: The figure consists of a diagram with 3 elements: "Physical Environment", A traced sketch of a user wearing a HoloLens 2 device while drawing an s-shape in mid-air, and "Spatial Search Query". "Physical environment" is connected to "Spatial Search Query" with an arrow, labelled "tracing". The sketch of the user is connected to "Spatial Search Query" with an arrow, labelled "sketching".
    Figure 1.a-1: A photo of an office table is depicted. In the background, a laptop, a flowerpot and a small 3D printer are positioned on the table. In the right half of the frame, a blue and white spatial sketch is augmented over the scene. It depicts the number 8 laid on its side or the infinity symbol. The application's result user interface is seen towards the top of the frame. The text field is filled with the word "Ring" and a grid of 4 by 5 results is shown below. Each result consists of a square thumbnail image and the title of the object. To the left of the image, a single white 3D model resides over a blue circle. It depicts two rings, joined on one side and rotated slightly inwards.  
    Figure 1.a-2: The photo shows an 3D-model of a mug with an additional sketched handle on the right side augmented over an office table. The user's hands are reaching into the frame from the bottom. To the left, a small portion of the search result interface of the application is seen.
    Figure 1.b: The figure consists of a diagram with 3 elements: "Physical Environment", A traced sketch of a user wearing a HoloLens 2 device while making a framing gesture and a blue square being formed by the hands, and "Textual Search Query". "Physical environment" is connected to "Textual Search Query" with an arrow, labelled "label extraction". The sketch of the user is connected to "Textual Search Query" with an arrow, labelled "search terms".
    Figure 1.b-1: The image shows a small black metal fan besides a laptop. Behind the fan, a flowerpot can be seen. In front of the fan, the two hands of the user are framing the fan. A blue rectangle is augmented over the scene, depicting this frame around the fan.
    Figure 1.b-2: The image shows the same scene as before (fan, pot and laptop on top of an office table. Additionally, various small objects are scattered across the table). The image shows the application's search interface with a list of labels from the image-based label extraction component as buttons. The term "spoke" is set in the input field of the search interface.}
    \caption{\system is a proof-of-concept implementation of in-situ spatial search. It enables users to search for future physical artifacts through spatial (a) or textual (b) queries, while coupling the search and previewing process to the users' physical context. \textbf{Spatial search queries} (a) rely on in-situ sketches, which may be drawn mid-air (1), be tracings of objects/features in the physical environment, or combine prior search results with sketched features (2). \textbf{Textual search queries} (b) rely on written or spoken terms known by the user. Alternatively, a user may use a photo-based label extraction to frame objects of interest (3) and receive labels to potentially use for searching (4).}
    \label{fig:teaser}
\end{teaserfigure}

\maketitle

\section{Introduction}
\label{sec:introduction}
Personal Fabrication is a powerful opportunity for technology enthusiasts and consumers alike: 
It empowers them to design and fabricate artifacts that are unrivaled in their degree of personalization and precision.
While technology enthusiasts may be compelled to invest time in the process of personal fabrication, consumers may generally care more about the result~\cite{hudsonUnderstandingNewcomers3D2016} and expect a low-friction workflow~\cite{stemasovRoadUbiquitousPersonal2021}.
To be able to include the whole spectrum of users to benefit from the progress of personal fabrication, future creation tools are increasingly addressing the needs of non-experts and laypeople~\cite{follmerCopyCADRemixingPhysical2010,savageMakersMarksPhysical2015,sunShrinkyKit3DPrinting2020}, which are often not able or not willing to express their needs with the precision an expert would.
This likely applies to the process of \emph{modeling} artifacts, but also to the process of \emph{searching and retrieving} the correct ones~\cite{stemasovEnablingUbiquitousPersonal2021}.
Through more affordable and robust hardware (e.g., 3D printers, laser cutters) and improved software workflows (e.g., modeling tools, slicers), personal fabrication is now in reach for a wider user base than ever before.\\

However, most established workflows used to define artifacts to fabricate (e.g., 3D modeling $\Rightarrow$ slicing $\Rightarrow$ 3D printing), require both learning and usage effort.
Such challenges, in part, explain the popularity and importance of open, crowd-based model repositories, such as \tv or \mmf~\cite{flathCopyTransformCombine2017,hudsonUnderstandingNewcomers3D2016,alcockBarriersUsingCustomizing2016}.
Model repositories provide users ready-to-print artifacts, making the interaction with the repository more related to shopping and less like established notions of modeling or designing artifacts~\cite{stemasovRoadUbiquitousPersonal2021}.
Model repositories are an appropriate alternative to modeling artifacts from the ground up, and will become even more viable as they grow in size and artifact diversity.
However, model repositories require feasible ways to be searched, to be a viable asset for users. 
Text-based search was adopted from established domains (e.g., file search), but largely misses the spatial nature and physical context of searching for (future) physical artifacts.
Similarly, query formulation and artifact previewing are disconnected from the physical context they are meant to be placed in later on.
When facing a requirement or challenge, users then are either required to invest effort in modeling artifacts or are required to translate their requirements to search terms and subsequently refine them.\\

We propose \system, a mixed-reality tool to search model repositories for 3D printing in-situ. 
It combines two types of search queries: textual and spatial.
Users may sketch coarse 3D shapes that are then used for geometry-based (i.e., spatial) searches.
They may likewise enter known search terms for textual searches.
To leverage the user's physical context, \system provides ways to support users in the task of defining and combining these two query types (Figure \ref{fig:teaser}): users are able to trace features of the environment for spatial searches, or may retrieve labels for objects in the environment for textual searches.
This bridges the disconnect between the users' physical environment and the search process.
\system was developed as an application for the Microsoft HoloLens 2\footnote{\url{https://www.microsoft.com/en-us/hololens}, retrieved on 12.12.2021} and is connected to a custom database of 3D-printable artifacts.
By leveraging additional, spatial modalities (i.e., sketches), \system allows users to omit the task of precisely defining search terms for artifact retrieval.
While some objects have clear and established terminology, more niche artifacts may require knowledge of the terminology used to describe them (i.e., a certain degree of domain knowledge). 
An example can be seen in Figure \ref{fig:teaser} a-1: a ring meant to be worn on two fingers simultaneously. 
Users may not necessarily know the term \emph{''double ring''}, but likely have an intuitive understanding of the \emph{geometry} they desire. 
Hence, they would be able to sketch the shape in a coarse fashion and use this sketch, combined with imprecise search terms (e.g., ''ring''), to find the desired artifact.
Similarly, users may have an intuitive understanding how an elongated vase with handles looks like, but may not know the term ''amphora'' used to \emph{precisely} denominate it.
By searching for geometry, users may receive results that fulfill their intended (functional) needs, without limiting themselves to objects \emph{originally} meant for their goal.
By conducting the sketching procedure in-situ (i.e., at the location of the future artifact), users benefit from previewing and referencing real-world features (e.g., by tracing them).
\system also enables users to get machine-generated labels and guesses after framing an object of interest (Figure \ref{fig:teaser} b).
This enables them to discover potentially unknown terms.
In turn, this helps finding similar objects, or objects meant to interact with the existing object.
By providing multiple ways (i.e., \emph{modalities}) to express and refine searches, \system aims to allow users to choose the path to a desired artifact that fits the task and the user best.
Users are also encouraged to iteratively refine their queries.
A fitting 3D model could be used as a base for a new spatial search query, by sketching new features onto it, as seen in Figure \ref{fig:teaser} a-2.
With \system, we take the perspective of 3D printing as a domain for physical artifact acquisition.
However, we believe that in-situ spatial search, and, by extension, \system, are applicable to the search for and acquisition of physical artifacts in general, regardless of the actual method to fabricate or acquire it.\\

\noindent The contributions of this work are as follows:
\begin{itemize}
    \item The \textbf{concept of in-situ spatial search for (physical) artifact retrieval}. It actively embeds both spatial input, textual input and the user's physical context in the search procedure. 
    \item \textbf{Proof-of-concept implementation of \system}, a Mixed-Reality application enabling in-situ spatial search of a model repository for 3D printing.
    \item \textbf{Exploration of the concept} of in-situ spatial search, using walkthroughs enacted with the \system prototype. The walkthroughs highlight novel ways for users to formulate and refine queries for future physical artifacts.
\end{itemize}

\section{Concept: In-situ Spatial Search}
\label{sec:concept}
    \begin{figure*}[ht!]
        \includegraphics[width=\linewidth]{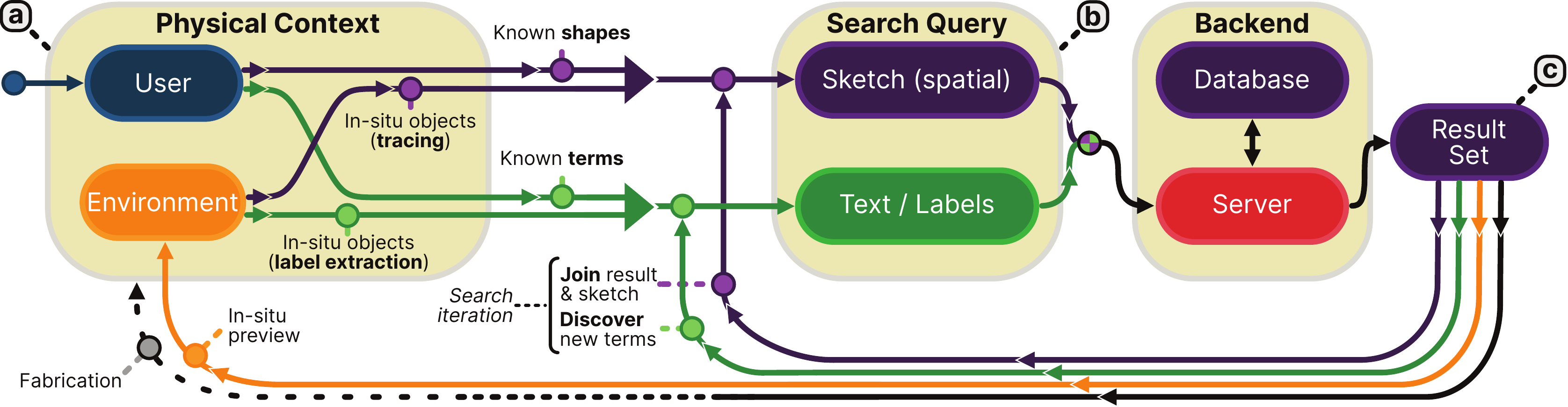}
        \caption{Conceptual process of in-situ spatial search, as demonstrated with \system. The users reside in the same physical context in which a future artifact may be fabricated for (a). They may search the repository with 3D sketches or known words, composing a search query (b). As the search happens \emph{in-situ}, users may also extract labels from the environment or trace existing objects. While refining their queries, users may learn new terms from the search results for textual searches and sketch new features onto the results. The query yields a result set (c) generated by the database server. Both query types (spatial, textual) can be combined, enabling multimodal search, while allowing users to be ambiguous in one or both modalities. }
        \label{fig:process}
        \Description{Figure 2: The figure depicts a diagram of the search process. Four core groups are labelled: the physical context (a), the search query (b), the backend, and the result set (c). Inside the physical context, there is a node representing the user and a node representing the environment. The search query contains a node named "sketch (spatial)" and a node named "text / labels". The user is connected to the spatial sketch through an arrow, labelled "known shapes". The environment is connected to the spatial sketch through an arrow labelled "In-situ objects (tracing)". The user is connected to the text with an arrow labelled "known terms". The environment is connected to the text with an arrow labelled "In-situ objects (label extraction)". The search query group's elements (sketch and text) are  connected to the server with an arrow. The server is connected to the database with an arrow. The Server is connected to the result set with an arrow. From the result set, 4 arrows point back to different locations. One points to the spatial sketch, labelled "join result and sketch". Another points to "text / labels", labelled "discover new terms". Another points to the environment, labelled "in-situ preview". The last point to the physical context, labelled "fabrication"}
    \end{figure*}

	With this section, we want to elaborate on the concept behind \system.
    \emph{In-situ} spatial search supports users with the \emph{transfer of features in their physical context to search queries}.
    This currently happens through users' mental efforts, domain knowledge, and through textual queries only.
    In-situ \emph{spatial} search also \emph{embraces the spatial nature of physical artifacts}, not only for result-presentation, but, where beneficial, also during the search process. 
    We argue that text-based search adds an unnecessary abstraction from a three-dimensional, spatial problem to a one-dimensional search term.
    Only in the final step, the retrieved artifact is transformed back to 3D by presenting a 3D object (either a digital, or a physical one) to the user. 
    This abstraction towards a 1D space (i.e., text) is helpful and powerful for expert users that know the (domain-)specific language and terminology.
    However, it creates an unnecessary burden for laypeople and novices who may be able to coarsely describe their 3D problem in-situ without necessarily knowing the appropriate and precise terms. \\
    
    The \textbf{in-situ} aspect of in-situ spatial search supports the transfer of features in the user's physical context to search queries. 
    Physical objects may be traced with (spatial) sketches, scanned (if the headset is capable of doing so), or be photographed to extract labels.
	These approaches transfer and translate context features (e.g., geometry) to search queries.
	With in-situ spatial search, more abstract aspects of context could also be available to use and embed in queries.
	Context may mean the room a user is in (e.g., the kitchen), the task the user is attempting to solve (e.g., attach an object to another), the general design of the room (e.g., ''modern''), or the color climate of the interior (e.g., earth tones to be matched by the future artifact).
	\system enables users to trace objects and apply photo-based label extraction.
	Apart from the use of the context's features for search queries, it is likewise used for previewing artifacts in context.
	This enables the user to estimate the form and function of the future artifact, while also allowing interactions like positioning and scaling to evaluate and improve the artifact prior to fabrication~\cite{stemasovMixMatchOmitting2020} or acquisition~\cite{leeConsumerCreatorHow2017} in general.\\

    The \textbf{spatial} aspect of in-situ spatial search embraces the spatial nature of physical artifacts during the search process.
    By embedding geometry and sketches in the search procedure, geometry receives an emphasis over terminology.
    If systems sidestep text and labels, functions and functionality ascribed to objects by contributors to an artifact database (e.g., users on \tv) can be ignored.
    The general topology of cups, pencil holders, vases, pots, and other cylindrical containers is comparable.
	However, this geometry of a bin is labeled differently by contributors, while all can be used for ''holding'' tasks (e.g., holding pencils).
	Not all artifacts \emph{suitable} for holding pencils can be found through the original label (e.g., \textit{''pencil''} or \textit{''pen''} holder).
	While geometry does not necessarily strictly \emph{define} function, it often is tightly coupled to it, particularly for static objects.
	Furthermore, we argue that users are likely to have at least a coarse understanding of their desired geometry or shape.
	Approximating it through a spatial sketch to find existing artifacts (to use or remix) may be a feasible procedure to retrieve a fitting one.
	This avoids the enforced transfer of spatial requirements (3D) to \emph{precise}, written keywords (1D), which is particularly relevant if a) fitting terminology is unknown or b) if the actual geometry of a future artifact is the most relevant aspect.\\

    We developed \system as a proof-of-concept prototype embodying our vision of in-situ spatial search.
    With the means of personal fabrication, the process of artifact acquisition can be approached from distinct directions: 1) design (e.g., 3D modeling or sculpting an artifact from the ground up) or 2) retrieval (e.g., searching for an artifact on \tv to fabricate).
    Both directions are equally viable to users focusing on the goal of ''attaining an artifact''.
    Usually, the design or modeling approach exhibits the highest complexity, but also enables sufficiently proficient users to achieve \emph{any} requirement they have~\cite{stemasovRoadUbiquitousPersonal2021}.
    However, designing artifacts requires knowledge and training, and therefore has a higher entry barrier. 
    In contrast, retrieving an artifact generally requires knowledge of the correct terminology to pass to a search system.
    Alternatively, users have to rely on recommender systems, or invest time in elaborate iterations and refinement.
    When dealing with high-level concepts (e.g., ''\textit{chair}'', ''\textit{vase}''), terms are generally established and there is little to no mismatch between the users' intentions and the understanding of the term by the artifact repository.
    The more specific and personal the requirements become, the more detailed and precise the search terms have to be to yield satisfying results.
    This makes the combination of multiple search modalities an appealing outlook.\\

    \system provides two main ways for searching: spatial queries (i.e., through sketching) and textual queries. 
    A diagram of the search process is depicted in figure \ref{fig:process}.
    Users are supported in the task of embedding their physical context in the queries: by tracing existing features for a spatial query, or by using label extraction to generate suggested search terms for a textual query.
    This is crucial, as any artifact users retrieve is meant to be fabricated or acquired otherwise -- ultimately interacting with the physical context~\cite{ashbrookAugmentedFabricationCombining2016}.
    In-situ search lowers the effort required to define searches and preview results in-context, as it omits \emph{transfers} between a location of (future) use (e.g., the table a vase will rest on) and a location of searching/design (e.g., a computer). 
    Both query types may also be created without involving the users’ physical context (e.g., by relying on known terminology or known geometry).
    This allows users to freely encode their existing wishes and knowledge, but demands a certain level of domain knowledge (for search terms) or at least a coarse understanding of the desired geometry.
    Through the use of in-situ spatial search procedures, a set of more specific advantages emerges for users: users may circumvent established terminology, ignore objects’ intended function and use a sketch's scale as a filter or a target scale. 
    In general, in-situ spatial search is meant to enable users to formulate and refine \emph{ambiguous} search queries in \emph{multiple modalities} to ideally retrieve finished, fitting designs from a model repository.\\

\section{Related Work}
\label{sec:relatedwork}
    The concept of in-situ spatial search draws inspiration from different research directions.
    \system is meant to be a low-effort, novice-friendly tool for personal fabrication, which skips established steps such as 3D modeling. %
    The development and usage of in-situ tools for design and fabrication is likewise a crucial direction related to \system.
    Similarly, novel methods for non-textual search and fabrication that do not necessarily follow established paradigms of (3D) modeling are fundamentally relevant to our work.
    Conceptually, the embedding and use of model repositories like \mmf or \tv is a relevant direction, as it emphasizes the benefits of crowd-sourced artifacts (i.e., artifacts that have already been designed and specified).
    
    \subsection{Novice-friendly Tools for Fabrication}
        Ongoing research efforts are being made to enable personal fabrication for a wider audience.
        This audience consists primarily of novices to design and fabrication (i.e., ''consumers'').
        As modeling artifacts is a complex task, attempts were made to simplify this modeling process, by focusing on fewer primitives~\cite{baudischKyub3DEditor2019}, 2D designs for 3D models~\cite{mccraeFlatFitFabInteractiveModeling2014}, or sketch-based modeling~\cite{saulSketchChairAllinoneChair2011}.
        Ballagas et al. explored voice input to generative models~\cite{ballagasExploringPervasiveMaking2019}, Lee et al. applied gestures to the task of furniture customization~\cite{leePosingActingInput2016}.
        CraftML by Yeh and Kim transferred 3D modeling to a declarative programming approach~\cite{yehCraftML3DModeling2018}.
        Apart from purely digital approaches, analog fabrication tools can be augmented and simplified, like MatchSticks~\cite{tianMatchSticksWoodworkingImprovisational2018} or Turn-by-Wire~\cite{tianTurnbyWireComputationallyMediated2019} by Tian et al.
        To support novices, automation and generative designs were used~\cite{shugrinaFabFormsCustomizable2015, liRobiotDesignTool2019,igarashiTeddySketchingInterface1999}, which infer the users' design intents based on often coarse input, such as sketches~\cite{kaziDreamSketchEarlyStage2017,saulSketchChairAllinoneChair2011,liSketch2CADSequentialCAD2020,johnsonSketchItMake2012}.
        
        All aforementioned tools can be seen as approaches to reduce effort, learning, and overcome challenges in personal fabrication.
        With in-situ spatial search and \system, we similarly aim to address novices to personal fabrication and design.
        However, we focus on outsourced design artifacts (i.e., 3D-printable models from experienced users uploaded to \tv or \mmf).
        This is a contrast to procedures inferring or generating geometry~\cite{liRobiotDesignTool2019,saulSketchChairAllinoneChair2011} or requiring users to model artifacts by aggregating primitives~\cite{yehCraftML3DModeling2018,baudischKyub3DEditor2019}.
        
    \subsection{In-situ Tools for Personal Fabrication}
        There is a disconnect between the location of artifact design and the artifact's future location~\cite{ashbrookAugmentedFabricationCombining2016,mahapatraBarriersEndUserDesigners2019}.
        This requires either assessment and transfer of requirements, or methods to compensate measurement errors~\cite{kimUnderstandingUncertaintyMeasurement2017}, which is a challenge that is not exclusive to novices.
        In-situ methods for modeling and fabrication enable users to preview designs within their future context.
        This was demonstrated by Peng et al.~\cite{pengRoMAInteractiveFabrication2018}, where design overlapped fabrication, or Yung et al.~\cite{yungPrinty3DInsituTangible2018}, where the target audience was children.
        In-situ tools relying on augmented or mixed reality also enable the augmentation and guidance of previously manual and analog fabrication approaches, such as 3D pen sculptures~\cite{yueWireDraw3DWire2017}, carved models~\cite{hattabInteractiveFabricationCSG2019,hattabRoughCarving3D2019}, or linkages~\cite{jeongMechanismPerfboardAugmented2018} by providing feedback during the process itself.
        They may also support remote collaboration and learning, as shown by Villanueva et al.~\cite{villanuevaRobotARAugmentedReality2021}.
        Mixed reality approaches, such as the works by Weichel et al.~\cite{weichelMixFabMixedrealityEnvironment2014} or Jeong et al.~\cite{jeongMechanismPerfboardAugmented2018} enable situated design approaches, while other tools focus on fabrication~\cite{zhuFusePrintDIY5D2016,pengRoMAInteractiveFabrication2018}, or remixing~\cite{stemasovMixMatchOmitting2020}.
        The usage and embedding of real-world artifacts as counterparts to future fabricated artifacts is likewise an aspect of in-situ tools.
        It enables the design of fitting mounts~\cite{zhuFusePrintDIY5D2016}, mechanisms actuating other objects~\cite{liRobiotDesignTool2019}, or previewing to-scale sketches in context~\cite{agrawalProtopiperPhysicallySketching2015}.
        
        Personal fabrication and any process that generates personal physical artifacts require grounding in the users' physical context.
        This is achieved through situating design or retrieval processes in this particular context.
        Mixed reality is an outstanding opportunity to achieve this type of interaction.
        \system aims to provide a similar benefit, and, for instance, enables users to trace real-world artifacts to use these features for their searches.
        However, our focus lies on search and retrieval, instead of modeling or design.
    
    \subsection{Alternative-modality Interfaces for Fabrication or Search}
        The combination of different input modalities enables users to bypass established metaphors for CAD (i.e., for the \emph{design} of artifacts) or information retrieval (i.e., for the \emph{search} for artifacts).
        This includes uses of sketches as a coarse input~\cite{saulSketchChairAllinoneChair2011,liSketch2CADSequentialCAD2020,lauSketchingPrototypingPersonalised2012}, but also gestures~\cite{leePosingActingInput2016,devStandUpUnderstandingBodypart2016,kimTangible3DHand2005} or speech~\cite{ballagasExploringPervasiveMaking2019}.
        Alternatively, tangible manipulation may be employed for input and provide appropriate feedback~\cite{teTADCADTangibleGestural2015,siuShapeCADAccessible3D2019} not found in most industrial CAD tools.
        Such ways to express and define geometry may, depending on the user, reduce the effort to design geometry~\cite{stemasovEphemeralFabricationExploring2022}.
        In contrast to the interfaces that focus on the \emph{design and generation} of artifacts, retrieval interfaces focus on the search for finished designs or parts~\cite{holzDataMimingInferring2011}.
        Fraser et al. presented ReMap, which embodies a prototype of multimodal search~\cite{fraserReMapLoweringBarrier2020}, an aspect we aimed to embrace with \system.
        On a more technical level, various approaches for the comparison of 3D models are present in the literature.
        These include approaches to align point sets~\cite{arunLeastSquaresFittingTwo1987,rusinkiewiczEfficientVariantsICP2001}, computation of object similarity~\cite{lamUsingMathematicalMorphology2011,kriegelEffectiveSimilaritySearch2003,bustosFeaturebasedSimilaritySearch2005}, and associated retrieval methods based on geometry~\cite{tangelderSurveyContentBased2004}.
        In particular, gesture-based systems, such as DataMiming by Holz and Wilson~\cite{holzDataMimingInferring2011} or sketch-based systems, such as the ones introduced by Pu et al.~\cite{pu2DSketchBasedUser2005} Enenhofer~\cite{enenhoferSpatialQuerybySketch1996} or Eitz et al.~\cite{eitzSketchbased3DShape2010} inspired \system.
        While DataMiming omitted visual feedback for the users and focused on gestures as input, it is conceptually similar to \system, as it uses a fairly natural way to describe artifacts, and offloads the precise definition or matching to other parties~\cite{holzDataMimingInferring2011}.
        \system does not omit visual feedback but considers it to be a crucial part of the search process. It also benefits from context features (tracing, label extraction) and embraces multiple modalities to give users more freedom and ambiguity.
        With platforms like Thangs~\cite{physnainc.Thangs3DModel2021} offering 3D-model based search for 3D models, the notion of searching with shapes over labels is already in reach for consumers, albeit in an ex-situ fashion.
        Giunchi et al presented a combination of sketching and speech for the retrieval of chair models in VR~\cite{giunchiMixingModalities3D2021b}.
        While conceptually similar, our focus lies on users' physical environments, instead of virtual ones. 
        This requires ways to transfer spatial requirements to searches.
        
        Such novel user interfaces for search or design are an opportunity to enable widespread access to personalized artifacts, by lowering the skill floor while retaining high expressivity for more proficient users.
        Novel modalities are applicable both to searching for existing artifacts~\cite{holzDataMimingInferring2011,sousaSketchbasedRetrievalDrawings2010} and generating new ones~\cite{ballagasExploringPervasiveMaking2019,leePosingActingInput2016,xuSketch2Scene2013}.
        \system can be considered to be a comparable approach to artifact retrieval, with an emphasis on coarse inputs that are -- ideally -- transferred to high-fidelity artifacts from a model repository.
        \system similarly aims to enable and support iterative search procedures, where users leverage their physical context and retrieved artifacts to formulate and refine searches.

    \subsection{Usage of Model Repositories}
        Model repositories (e.g., \tv) and their associated communities of makers are an important component of today's personal fabrication landscape.
        Repositories have been leveraged as a model source for remixing tools, such as Grafter by Roumen et al.~\cite{roumenGrafterRemixing3DPrinted2018}, or Mix\&Match by Stemasov et al.~\cite{stemasovMixMatchOmitting2020}.
        They have also been the subject of studies and investigations, generating novel frameworks like PARTs by Hofmann et al.~\cite{hofmannGreaterSumIts2018} that aim to increase re-use of design effort.
        Likewise, systematic analyses of repositories like \tv and their users unveiled challenges in terms of artifact use and customization by novices~\cite{alcockBarriersUsingCustomizing2016,hudsonUnderstandingNewcomers3D2016}.
        Works by Flath et al.~\cite{flathCopyTransformCombine2017} or Kyriakou et al.~\cite{kyriakouKnowledgeReuseCustomization2017} evaluated patterns and degrees of knowledge and design reuse.
        Domain-specific usage, such as the exchange of assistive technologies~\cite{buehlerSharingCaringAssistive2015} is likewise a field benefiting from a combination of design sharing, re-use, and customization.
        
        We agree with and embrace the notion that crowdsourced models or artifacts are crucial for personal fabrication.
        Notably, any storefront -- on- and offline -- provides users a similar experience, with varying degrees of potential personalization.
        When considering stores to be an alternative to personal fabrication (currently chosen by a majority of the population~\cite{stemasovRoadUbiquitousPersonal2021}), approaches relying on searching and customizing existing designs can be considered to be a viable path to take.
        We approach this aspect with an emphasis on search and how it is applied to the task of finding future artifacts for one's own, personal, physical context.

\section{Proof of Concept Implementation}
\label{sec:implementation}
    The following sections describe the development of \system as proof-of-concept implementation.
    The system consists of 3 components: the application for Microsoft's HoloLens 2, a server component connected to the database, and a data-gathering tool which was used to fill the database with models gathered from \tv and \mmf.
    We report our approach and development process openly for replicability.
    
    \subsection{HoloLens 2 Application}\label{sec:hololens}
        We implemented the user-facing part of \system with Microsoft's HoloLens 2 using Unity\footnote{\url{https://unity.com/products/unity-platform}, retrieved on 16.12.2021} 2019.4.6f1.
        Version 2.4 of the Mixed Reality Toolkit (MRTK\footnote{\url{https://microsoft.github.io/MixedRealityToolkit-Unity}, retrieved on 26.8.2021}) was used to implement the user interface and most of the interactions with it.
        Networking was implemented with the RestClient\footnote{\url{https://github.com/proyecto26/RestClient}, retrieved on 15.12.2021} library for Unity.
        
        \begin{figure}[hb!]
            \includegraphics[width=\minof{\columnwidth}{0.65\textwidth}]{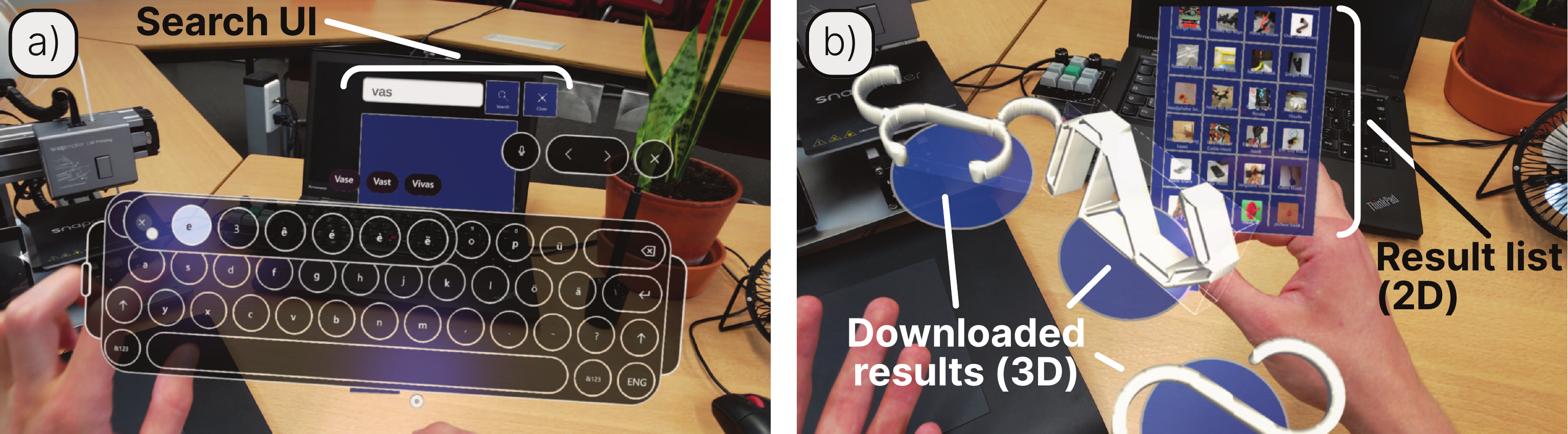}
            \caption{The textual search interface, along with the keyboard for text input (a). After submitting the search query, a user is presented with a 2D list of results and thumbnails (b, right). Downloaded objects are rendered in 3D and are attached to the user's left hand (b, left).}
            \Description{Figure 3.a: The photo depicts a view towards an office table. In the bottom half of the frame, a keyboard is augmented over the view. It is semi-transparent, black and has round keys. The user's left hand is reaching into the frame from the bottom left corner and is currently typing the letter 'e'. Three word suggestions are shown by the keyboard: 'vase', 'vast', 'vivas'. Behind the keyboard, the search user interface of the application can be seen. The search text field currently contains the letters 'vas'. To the right of it, a search button is seen, labelled 'search' and showing a magnifying glass icon. To the right of it, a button labelled 'clear' is seen, showing an 'x'-icon. Below this row of elements, a plain blue panel is seen, with half of it being covered by the keyboard which is augmented closer to the viewer.
            Figure 3.b: The photo depicts a view of the same office table. The user's hands are reaching into the frame: the left hand is rotated with the palm towards the viewer. The right hand is making a pinch gesture (Thumb and index finger are connected, the others are spread apart slightly). Three blue circles are augmented over the image, arranged around the left hand. Above each circle, a white 3D-model of a hook is visible. The first hook is rounded and shows one oval mounting clamp and two hooks protruding from it. The second hook is comprised of straight shapes and lines. An incision to allow the hook to be mounted to a table or a shelf is seen. The third hook is a plain s-shaped hook. To the right of the 3 hooks, the application result interface is seen. It is a grid of 6 by 4 search results. Each result is arranged in a square shape, with a square thumbnail above the object's name.}
            \label{fig:search-and-carousel}
        \end{figure}
        
        The search interface is initially attached to the user's right hand.
        If needed, the user can grab and position the panel statically in space (Fig. \ref{fig:search-and-carousel}a).
        The search interface can be used to enter search terms by text or speech.
        After submitting a query, the interface loads a scrollable grid of results (Fig. \ref{fig:search-and-carousel}b).
        The user can select results that appear promising, which are then downloaded in the background.
        As soon as the download and instantiation are completed, the model appears attached to the user's left hand, forming a palette where the user may gather up to 5 models to compare and evaluate (Fig. \ref{fig:search-and-carousel}b).
        Each artifact loaded into the palette can be grabbed and positioned in the space around the user.
        The objects can then be repositioned, rotated, and scaled freely, which enables an in-situ preview of them.
        
         \begin{figure}[hb]
            \includegraphics[width=\minof{\columnwidth}{0.65\textwidth}]{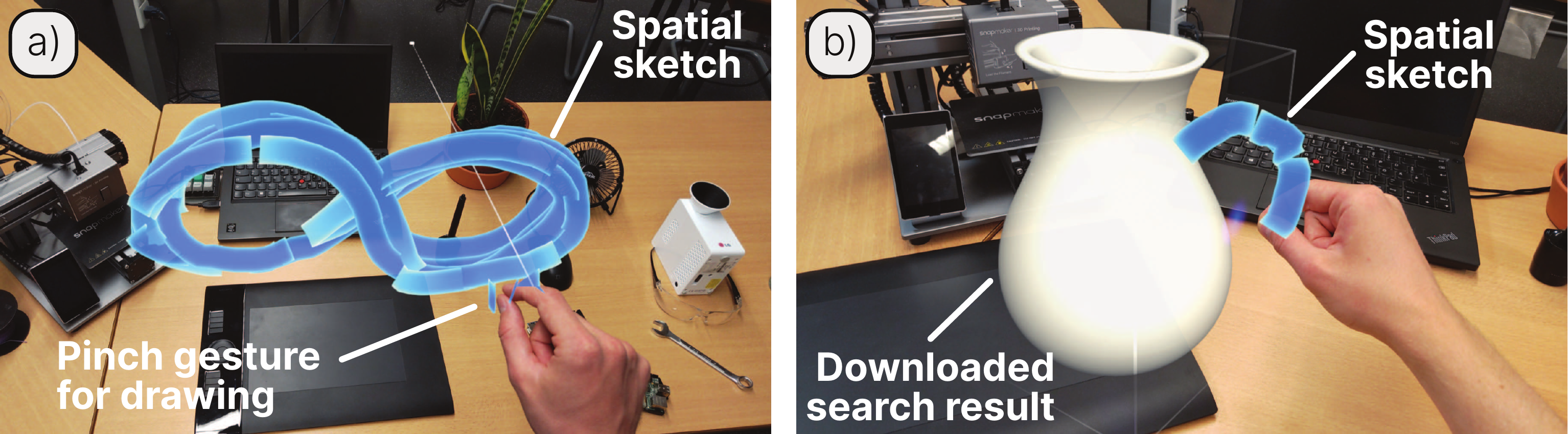}
            \caption{The sketching mode of \system. The users are free to sketch queries in the space around them with a pinch gesture (a). To iteratively refine spatial search queries, users may also sketch additions onto an object retrieved from the repository (b) and submit this combination as a new query. }
            \Description{Figure 4.a: The photo depicts a top-down view of an office table. Various items are scattered across it, such as a laptop, a small 3D printer, a flowerpot and a black metal fan. A blue spatial sketch is augmented front and center over the scenery. The sketch has the shape of an 8 laid flat (i.e., an infinity symbol). The user's right hand is reaching into the frame from the bottom edge. The user is making a pinch gesture (thumb & index finger).
            Figure 4.b: The same office table scene is seen from a flatter angle. A large white vase is augmented over the table. The vase is plain and rounded, with a wide neck and slightly wider body. The user's hand is reaching into the frame from the bottom right corner of the image. The user is making a pinch gesture and is thereby sketching an additional feature onto the vase: a handle, which starts right below the vase's neck and is not completely finished. The sketched-on handle is rendered in a blue color.
}
            \label{fig:sketch-search}
        \end{figure}
        
        The sketch search component can be activated by tapping on a button below the search interface.
        After activating the sketch mode, the users can draw in the space around them by pinching their thumb and index finger (Fig. \ref{fig:sketch-search}a). 
        The sketch can then be submitted as a query.
        To ensure appropriate results, sketch-based search currently requires the users to enter at least a coarse search term. 
        This is mainly used to refine the search to receive faster results and can be removed in the future with either better (backend) hardware or a more optimized implementation of the matching algorithm. 
        The users may also sketch onto downloaded and positioned meshes, to extend them and use them as a new query (Fig. \ref{fig:sketch-search}b).
        This feature also enables users to position multiple models they retrieved from the repository and connect them through their sketch, which is then treated as a single spatial query by the server.
        Thereby, users do not necessarily have to sketch entire objects, but may start with an object and add features they deem missing from it (e.g., adding handles to a vase).
        
        \begin{figure}[ht]
            \includegraphics[width=\minof{\columnwidth}{0.65\textwidth}]{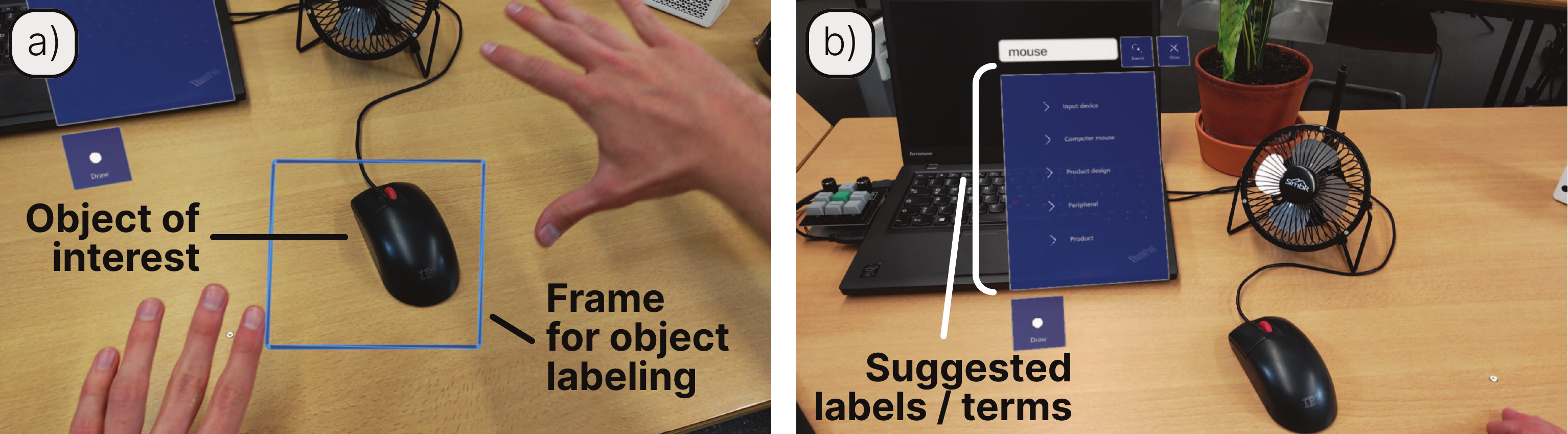}
            \caption{The image-based label extraction of \system: The user activates a framing mode through a gesture, encloses the artifact to be classified in the frame, and confirms the selection via dwell-time (a). The ''best guess'' is pre-filled into the search field and additional suggested keywords are listed below (b).}
            \label{fig:photo-search}
            \Description{Figure 5.a:    A top-down view of a table is seen. The user's hands reach from the bottom edge into the frame. The image shows how the user frames a black computer mouse with a red scroll wheel with both hands. The hands are positioned at 2 corners of a blue rectangle framing the mouse.              
            Figure 5.b: The image shows the application's search interface augmented and hovering over an office table. In the interface, the results from the image-based label extraction are listed as buttons. The term "mouse" is filled in the input field of the search interface
}
        \end{figure}
        
        The image-based label retrieval extracts possible terms for textual queries based on an image captured by the headset.
        This component of \system is fundamentally a mapping from a 2D image to a search term (1D).
        It is enabled by connecting and holding the index fingers and thumbs of both hands for one second (dwell time) in the field of view of the device.
        This activates a ''framing mode'' (Fig. \ref{fig:photo-search}a).
        Afterwards, the users can drag their hands apart, to form a frame around the object they want to use as a query (Fig. \ref{fig:photo-search}a).
        The captured image is used as a query to the Google Cloud Vision API\footnote{\url{https://cloud.google.com/vision/}, retrieved on 11.12.2021}, which retrieves suggested labels based on the image's contents. 

    \subsection{Data Collection}\label{sec:crawler}
        While \tv and \mmf provide crucial infrastructure for makers, they are -- by design -- focused on established ways to be searched: textual queries along with sorting and filtering options.
        Notably, most filters focus on tags and categories (i.e., more abstract terminology).
        Sorting similarly focuses on non-spatial dimensions, like popularity or (textual) relevance.
        We chose \tv and \mmf as data sources for \system.
        \tv can be considered the de-facto standard for model repositories for 3D printing, while \mmf aims to curate the models offered on the platform.
        However, neither of these platforms offers ways for geometry-based searches\footnote{Implicitly, users may search for specific measurements or labels and hope to find results through textual search}.
        Thangs\footnote{\url{https://thangs.com/}, retrieved on 16.12.2021} allows users to upload 3D models as queries, but offers no way to define them (either coarse sketches or precise models), while also functioning ex-situ only~\cite{physnainc.Thangs3DModel2021}.
        To comply with the platforms' API guidelines and to enable our novel functionalities without straining public infrastructure, we chose to duplicate a subset of their libraries.
        We chose high waiting times between all requests, to avoid putting unnecessary strain on the servers of \tv and \mmf. 
        This led to timespans of multiple days to gather batches of 500 models.
        The keywords used were basic mechanical artifacts popular in 3D printing communities like hooks, and decorative artifacts like vases or figures.
        Furthermore, the most popular artifacts on each of the repositories were gathered.
        Data like the original source, the designer, the licenses used were all transferred to our database subset.
        
        The entire dataset we gathered consists of  4118  objects in total. 
        2717 objects consist of one part only, while the remaining ones have 2 or more files (e.g., .stl) associated with them. 
        2356 objects out of 4118 provide mesh metadata suitable for our implementation of spatial search queries.
        
    \subsection{Postprocessing}\label{sec:postprocessing}
        Postprocessing of the metadata and the meshes themselves was necessary after each iteration of data gathering.
        The tool used for mesh handling was the trimesh library for Python by Dawson--Haggerty et al. \cite{dawson-haggertyTrimesh2020}.
        Some meshes were corrupted, and were for instance missing correctly calculated normals or did not describe a volume.
        We did not apply any ''opinionated'' automated mesh repair methods\footnote{e.g., ones that fill holes larger than 1 triangle in size or remove self-intersecting or degenerate triangles, which are present in repositories like \tv~\cite{zhouThingi10KDataset102016}}, due to the risk of corrupting the original geometry.
        The data sources \tv and \mmf both employ the concepts of \emph{categories} and \emph{tags}.
        While tags are largely unmoderated and can be freely chosen by the users, available categories are specified by the repository administration.
        The categories used by \tv and \mmf are comparable, but not identical.
        Categories were therefore grouped and matched into new categories used by our database. 
        
        To improve the text- and voice-based search, fields like the artifact name, description, or tags were subject to postprocessing with NLTK\footnote{\url{https://www.nltk.org/}, retrieved on 1.12.2021} (stemming, lemmatization, and stopword removal).
        To enable geometry-based searches, spatial metadata of the meshes was also established during this step.
        As detailed registrations and comparisons of meshes are computationally expensive, we implemented methods to pre-filter the result.
        We calculated two ratios based on the ordered dimensions of the bounding box to filter meshes based on their proportions.
        The voxel representation used for aligning and matching sketches to objects was computed at this step and serialized to the database, to avoid re-computing it on each search request.
        
    \subsection{Server Implementation}\label{sec:server}
        All server-side software was written in Python.
        The server component was written using flask\footnote{\url{https://flask.palletsprojects.com/}, retrieved on 26.8.2021}. 
        The flask application provides a REST (representational state transfer) API for the HoloLens client: searches by text, searches by sketch, download of metadata, files, and thumbnails.
        For the database component, MongoDB\footnote{\url{https://www.mongodb.com/}, retrieved on 22.8.2021} 4.4.0 was used, due to its flexible data model and its document-based approach.
        To accelerate queries, our database instance employs multiple indexes.
        Apart from indexing the ID fields, which accelerates retrieval of object details and files, a textual index and a 2D spatial index are used.
        This ideally accelerates searches submitted through their respective index.
        For queries ultimately converted to text, the database uses a text index\footnote{\url{https://docs.mongodb.com/manual/core/index-text/}, retrieved on 1.8.2021}, based on the title, description, tag, and category fields.
        The fields are weighted based on their specificity (with categories being the least and names being the most specific).
        Searches passed to this index omit stopwords and employ stemming and lemmatization for the English language\footnote{Notably, not all gathered artifacts are named and described in English. This is an additional argument in favor of non-textual queries.}.
        A second index configured for the database is a 2D spatial index\footnote{\url{https://docs.mongodb.com/manual/core/2d/}, retrieved on 1.8.2021}.
        We leverage the accelerated query time for 2D-data we derive from the meshes' object-aligned bounding box (OABB).
        The ratios between height/width and width/depth are used to pre-filter models based on their \emph{proportions}.\\ 
        
        Each search request is initially treated as a textual search, but gets by highly imprecise and general terms (e.g., ''object'').
        This enables pre-filtering the results to amounts manageable by the server for computationally expensive operations and manageable by the user to get an appropriate overview.
        After having reduced the potential search result set, the server filters the results based on the previously introduced metric of OABB ratios.
        This set of results that match based on the OABB ratio is intersected with the set retrieved by the textual search.
        With this selection, it is now possible to calculate per-voxel overlap, which is the server's ranking/scoring function.
        The stored voxel representations are already size-normalized (i.e., their largest extent is set to 100 and the other extents are scaled proportionally to match).
        The incoming sketch is similarly normalized in scale.
        However, the original extents are retained, if the resulting model is meant to be scaled to match it.
        The sketch is then voxelized.
        The resolution of the voxelization is likewise defined by the model's largest dimension, as the voxelization pitch was chosen to yield 20 voxels across the largest dimension.
        The voxelized meshes are then used to run the ICP algorithm (iterative closest point~\cite{zhangIterativePointMatching1994}).
        The ICP procedure aligns the sketch to the repository model, as the latter usually has an orientation that follows the principal axes of inertia.
        The starting parameters for the ICP procedure are chosen based on the inertia of the mesh.
        The voxelized sketch is then rotated and moved with the calculated transformation.
        Due to the previously conducted size-normalization, no scaling is involved in this step.
        With the 2 models aligned and with maximal overlap, the amount (i.e., the number of voxels) of overlap is counted.
        Based on the resulting overlap, we calculate multiple similarity metrics: normalized by the number of voxels in the sketch, normalized by the number of voxels in the repository model, and an average of these two normalized values.
        To accommodate for different volumes (e.g., sketches being entirely wrapped by the repository model), the average of the two normalized values is used as the core ranking for the search results.

\section{Application Walkthroughs}
\label{sec:scenarios}
    The following scenarios were enacted with the \system prototype and present brief, self-contained examples of actual interactions with the system.
    The search results seen in the examples are based on our custom dataset and are therefore a \emph{subset} of models available on \tv or \mmf.
    However, due to a sizeable number of different objects of different categories, the user interacts with the dataset and the search results in a similar fashion to the potential interaction happening with a more extensive database.
    Furthermore, the sketch search is not ideally calculating similarities, but fails in specific edge cases (e.g., completely flat sketches, or when the user's sketch has entirely different proportions than any fitting object in the database).
    It does, however, generally retrieve a set of results where fitting or similar artifacts can be found among the first ones and iterated with.
    With the walkthroughs, we want to emphasize different ways in which in-situ spatial search enables users to ambiguously and iteratively approach physical artifact retrieval.
    
    \subsection{Scenario 1: Bypassing Terminology}
        \system supports users by affording them a degree of \emph{Term}--\emph{Abstraction}: not having to know or choose the terminology of a domain.
        This is enabled through the use of photo-based label extraction (2D $\rightarrow$ 1D) and sketch-based search (3D $\rightarrow$ 3D).
        
        \begin{figure}[b] 
            \includegraphics[width=\minof{\columnwidth}{0.65\textwidth}]{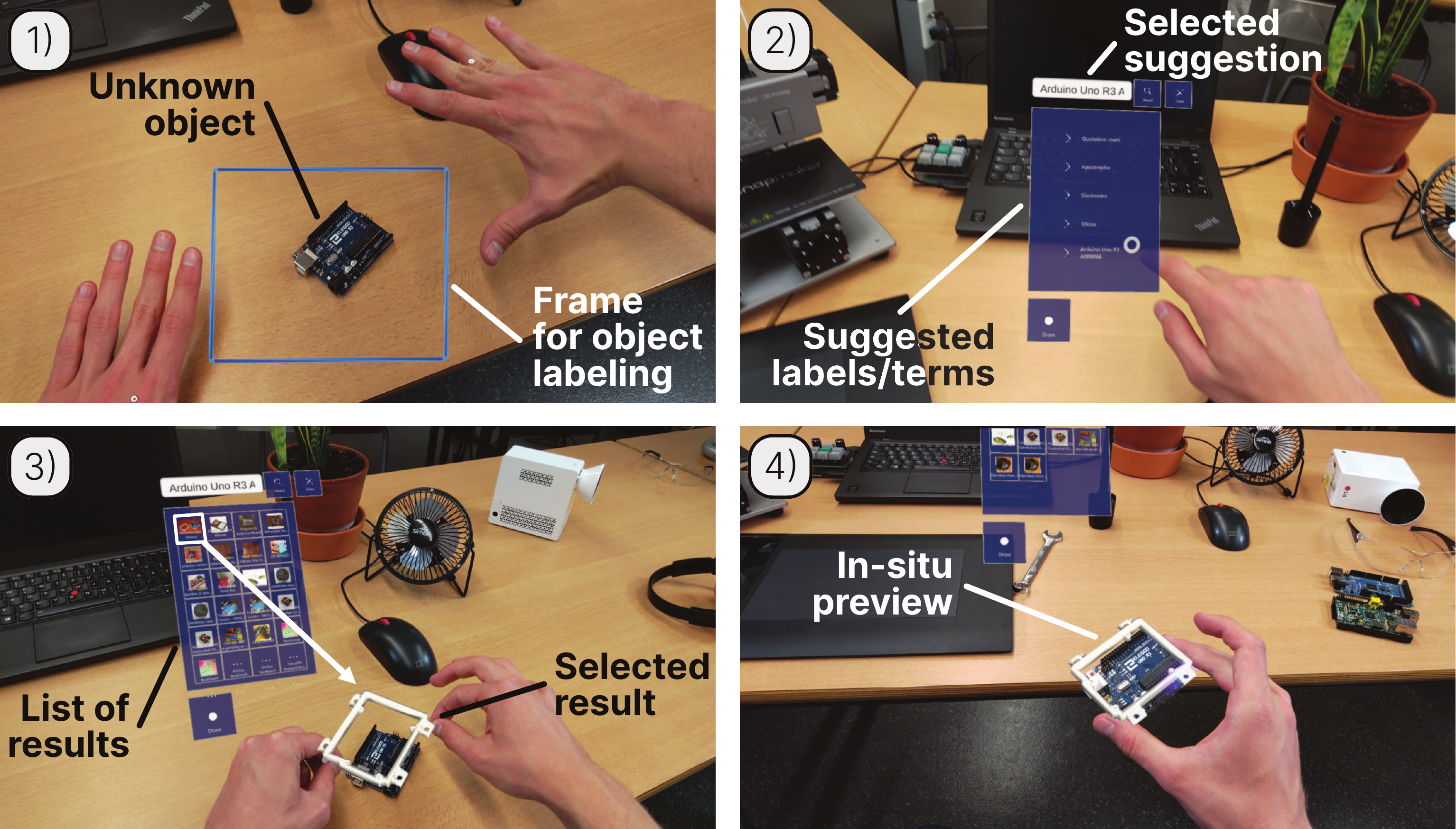}
            \caption{\textbf{Term-Abstraction}: Finding artifacts when terminology is unknown. 1) Framing the object to be searched for; 2) using one of the classified labels; 3) search results and scaling of a selected one; 4) previewing the result in-situ.}
            \Description{Figure 6.1: A top-down view of a table is seen. The image shows how the user frames an Arduino (microcontroller) placed on an office table with both hands. The hands are positioned at 2 corners of a blue rectangle framing the microcontroller. Figure 6.2: The image shows the application's search interface over an office table. The results of the label extraction are listed as buttons in the search interface. The input field of the search interface is set to "Arduino Uno R3". The user's right hand is reaching into the frame and is making a pointing gesture. The user's right index finger is hovering over a result with the label "Arduino Uno R3".
            Figure 6.3: The image shows the application's search interface filled with a grid of 6 by 4 results. "Arduino Uno R3" is filled in the input field. Both hands of the user are making a pinch gesture on the opposite side of a 3D-model of an arduino mounting case. The model is rendered above the scene in white and is a rectangular frame with 4 holes intended for screws.
            Figure 6.4: The office table is now farther back. The user's right hand is reaching into the frame and is holding the microcontroller between thumb and middle finger. The white mounting case is augmented over it and roughly matches its size. Farther back, the search interface of the application is floating over the table.}
            \label{fig:scenario-arduino}
        \end{figure}
        
        The first example can be seen in Figure \ref{fig:scenario-arduino}.
        A microcontroller (Arduino Uno), is in a user's vicinity.
        The name ''Arduino'' does not appear on the device, as it is made by an alternative manufacturer -- however, this would be an ideal term to use in a search.
        Not knowing what it exactly is (i.e., what to search for specifically), the user may want to find a case to enclose or mount the exposed microcontroller.
        He is able to use the photo-based label extraction component to frame the device (Fig. \ref{fig:scenario-arduino}-1) and is suggested the term ''Arduino'' among others (Fig. \ref{fig:scenario-arduino}-2).
        Executing this textual query yields a set of related models for 3D printing (Fig. \ref{fig:scenario-arduino}-3).
        This includes cases, mounts, but also objects that require an Arduino to provide interactivity.
        The models can be downloaded and previewed in situ (Fig. \ref{fig:scenario-arduino}-4), for the user to verify the appeal, and, to a degree, the fit of the model.

        \begin{figure}[htb] 
            \includegraphics[width=\minof{\columnwidth}{0.65\textwidth}]{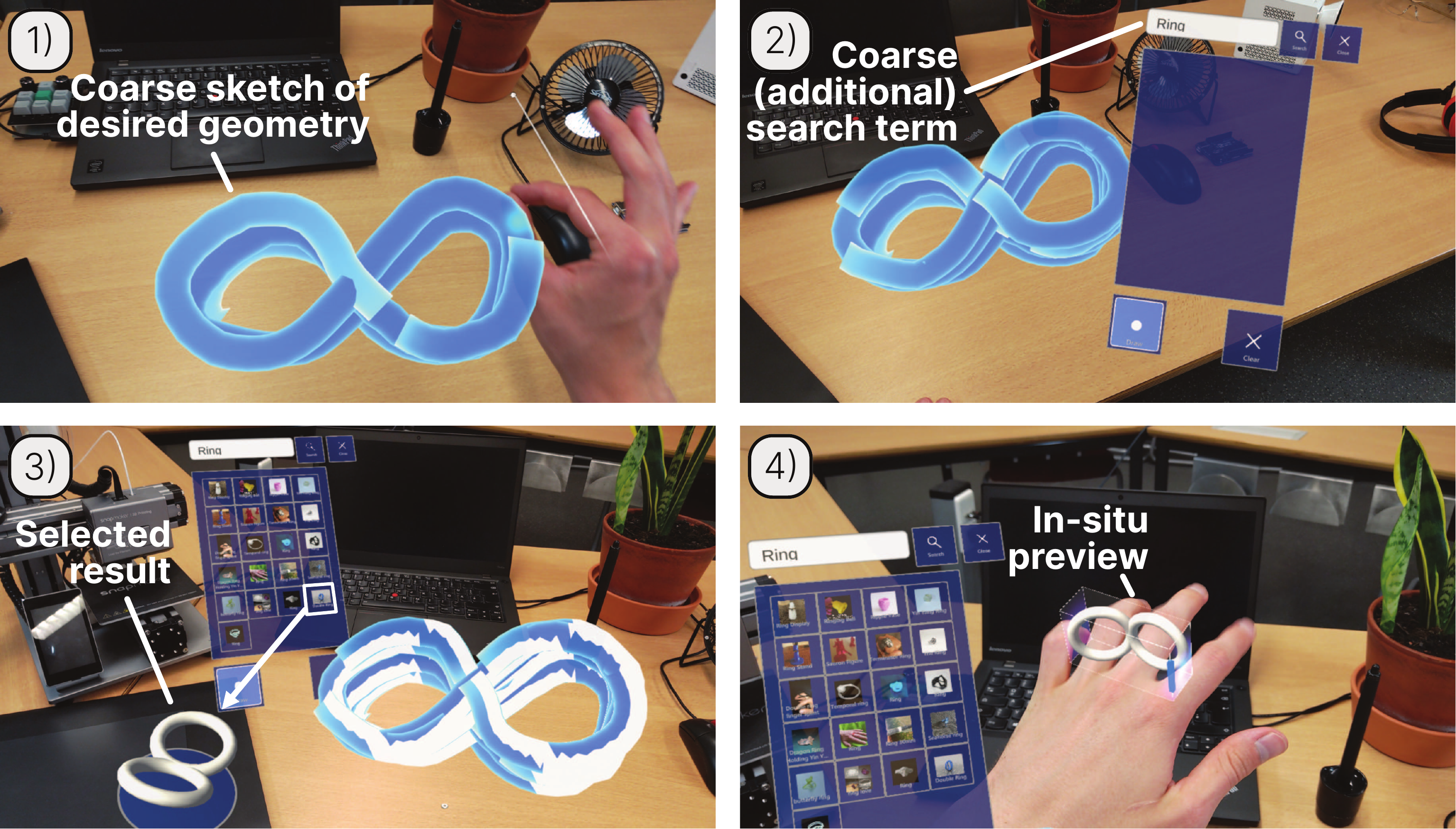}
            \caption{\textbf{Term-Abstraction}: bypassing domain terminology of jewelry. 1) sketching the coarse geometry of a double ring; 2) entering a coarse textual search term; 3) analyzing the list of (2D) results (input sketch and one fitting 3D result are visible); 4) previewing the (3D) object on the body.}
            \Description{Figure 7.1:    A photo of a table is visible. On the table, a laptop, a flowerpot and a fan are positioned. In the foreground, a 3D sketch in the shape of the infinity symbol is being drawn by the user. The user's right hand reaches into the frame and is making a pinch gesture (connecting the thumb and the index finger).
             Figure 7.2:    The system's search UI is seen towards the right side of the photo. the term "ring" was entered as a search term and an empty blue panel is floating below it. Below that, a button labelled 'draw' is highlighted. A button 'clear' is to the right of it.
             Figure 7.3:    The system's search UI is seen towards the top half of the photo. the term "ring" is entered as a search term and a grid of 6 by 4 results is visible (names and thumbnail images). To the left, a white 3D model of a ring shaped like the infinity symbol is augmented over the scene. The two rings it consists of are tilted slightly inwards. To the right, the initially drawn sketch is visible.
             Figure 7.4:    The system's search UI is seen towards the left side of the photo. the term "ring" was entered as a search term and a grid of 5 by 4 results is visible (names and thumbnail images). The user's left hand is reaching into the frame. The ring from the previous frame is augmented over the hand, with each part-ring residing over a finger (middle finger and ring finger).
}
            \label{fig:scenario-double-ring}
        \end{figure}
    
        Another example is depicted in Figure \ref{fig:scenario-double-ring}, where the user is interested in finding a double-ring\footnote{An example is \href{https://www.thingiverse.com/thing:2672476}{Double Ring JVCR (22 mm x 2) by cepera3000} on \tv, retrieved on 15.12.2021} which is a piece of jewelry worn on two fingers, instead of one.
        Most users interested in such a construct likely have an intuitive idea of the \emph{geometry} of the object: two similarly sized rings, possibly connected through a plate or other decorative element.
        However, they may not know the \emph{ideal term} to find such objects in a repository.
        A user can then sketch this geometry (2 rings and, if needed, a connecting plate -- Fig. \ref{fig:scenario-double-ring}-1), and use this in conjunction with a coarse search term (e.g., ''ring'', which would not retrieve the desired double ring as a high ranking result -- Fig. \ref{fig:scenario-double-ring}-2).
        While actual double rings do not rank first in the \system search, they are among the first results presented to the user (Fig. \ref{fig:scenario-double-ring}-3).
        Lastly, the user may preview the ring in–situ, which may not guarantee a perfect fit, but at the very least indicates fit and aesthetics (Fig. \ref{fig:scenario-double-ring}-4).

    \subsection{Scenario 2: Iterative Refinement}
        We do not expect users to formulate ''ideal'' queries that lead to the desired results in one single step.
        Refinement in textual searches usually happens by browsing initial results and applying filters.
        Alternatively, users may discover new terms to search for by browsing through item titles or descriptions, thereby acquiring a degree of domain knowledge.
        \system enables both a textual and a spatial query refinement process.
        
        \begin{figure}[htb]
            \includegraphics[width=\minof{\columnwidth}{0.65\textwidth}]{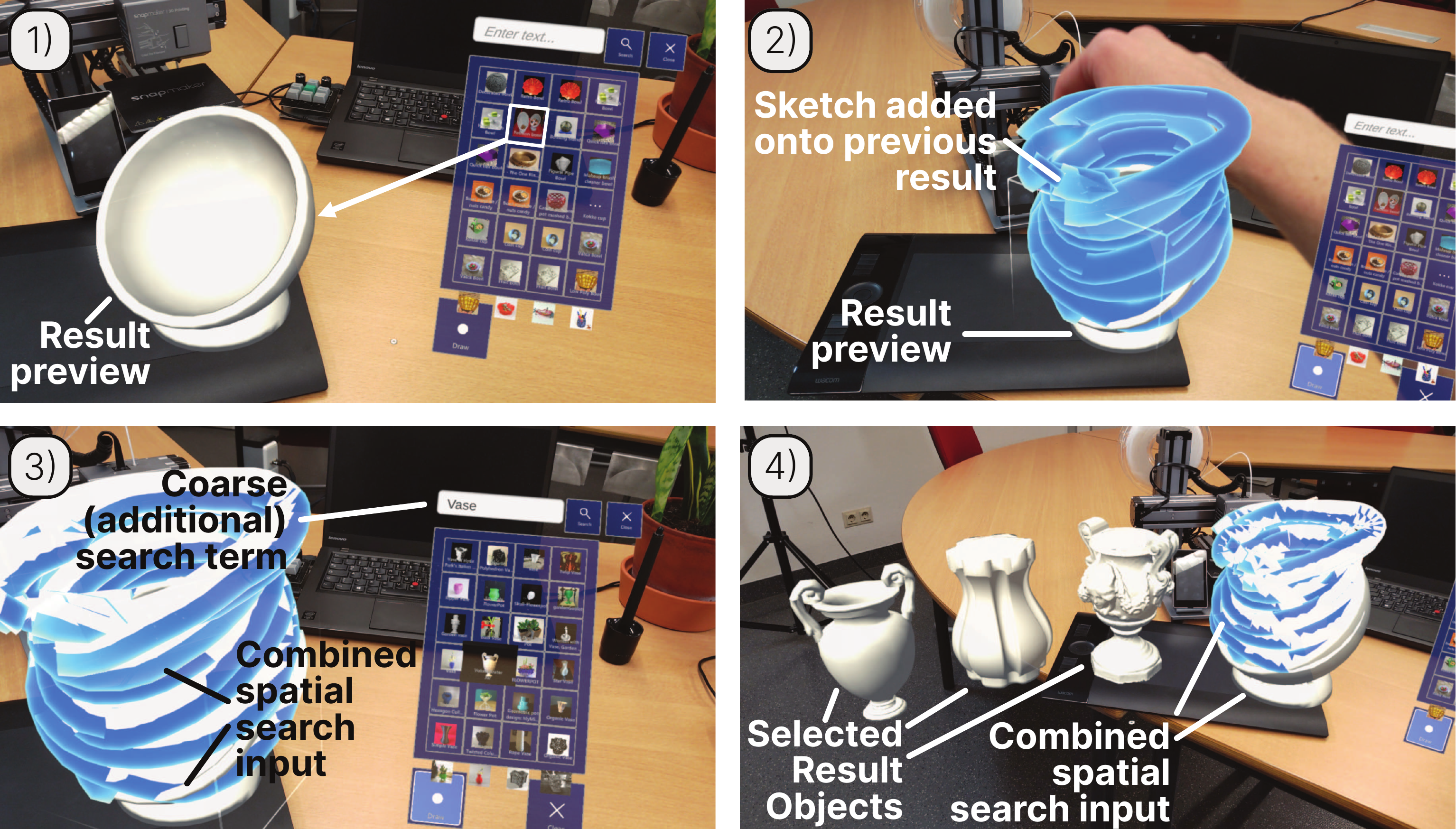}
            \caption{\textbf{Iteration (spatial):} Sketching onto retrieved artifacts to iterate on the (spatial) search query. 1) the initial bowl retrieved from the repository; 2) Sketching onto it to elongate its shape; 3) submitting the sketch search with a new term; 4) a set of results for the combination of sketch and model.}
            \Description{Figure 8.1:    The system's search UI is seen towards the right side of the photo. A grid of 6 by 4 results is visible (names and thumbnail images). To the left, a white bowl is augmented over the table. It consists of a base and a half-sphere, which is tilted towards the viewer.
             Figure 8.2:    The previously mentioned bowl is covered by a spiral-like blue sketch which elongates the half-sphere part towards the top. The user's hand is reaching into the frame, making a pinch gesture to draw the sketch in 3D space above the bowl.
             Figure 8.3:    The combination of sketch and bowl is seen in the right half of the photo. The table scene is in the background and the system's search UI is seen towards the right side of the photo. the term "vase" was entered as a search term and a grid of 6 by 4 results is visible (names and thumbnail images)
            Figure 8.4:    A zoomed out photo of the table scene is visible, with 4 vase- or pot-like objects augmented over the scene. The vase on the left is wide and has 2 decorative handles on the side. The second vase is wider at the bottom and has a wide neck. It has multiple vertical creases across its perimeter. The third vase has the shape of a chalice and has floral decorations across its perimeter. The last vase is the previously mentioned combination of a bowl and the sketch that elongates it.}
            \label{fig:scenario-pots}
        \end{figure}
        
        An example can be seen in Figure \ref{fig:scenario-pots}.
        The user initially searches for bowls to use for decoration (Fig. \ref{fig:scenario-pots}-1).
        As she notices that she would prefer ones that are generally taller, she can either take one of the pots retrieved from the repository and scale it in the Y direction, or she can enable the sketching mode and add onto the downloaded pot to enlarge it (Fig. \ref{fig:scenario-pots}-2).
        After having changed the pot's proportions this way, she may submit the combination of previous result and sketch, along with a coarse search term  (Fig. \ref{fig:scenario-pots}-3) as a new query.
        Lastly, she may preview the results in-situ and apply other manipulations such as scaling to the (3D) results  (Fig. \ref{fig:scenario-pots}-4).
        While the initial query is formulated without leveraging in-situ features, the query refinement may happen through referencing the physical context (e.g., comparing proportions of the results with existing decoration and altering it accordingly).

    \subsection{Scenario 3: Inspiration}
        Photo-based label extraction can be employed to provide inspiration and related items to the user. 
        By applying this label extraction, users transfer aspects of their physical context to textual queries. 
        This is similar to features found on platforms like Pinterest~\cite{wagnerPinterestWillNow2017} or shopping interfaces like IKEA's smartphone app~\cite{stinsonIkeaARApp2018},
        The items retrieved do not necessarily have to be the same artifact, but be similar to it, in terms of use-cases or domain. 
        For a phone, the search may retrieve phone mounts or cases, which do not exhibit an identical geometry, but are meant to complement the geometry of the artifact.
        
        \begin{figure}[htb]
            \includegraphics[width=\minof{\columnwidth}{0.65\textwidth}]{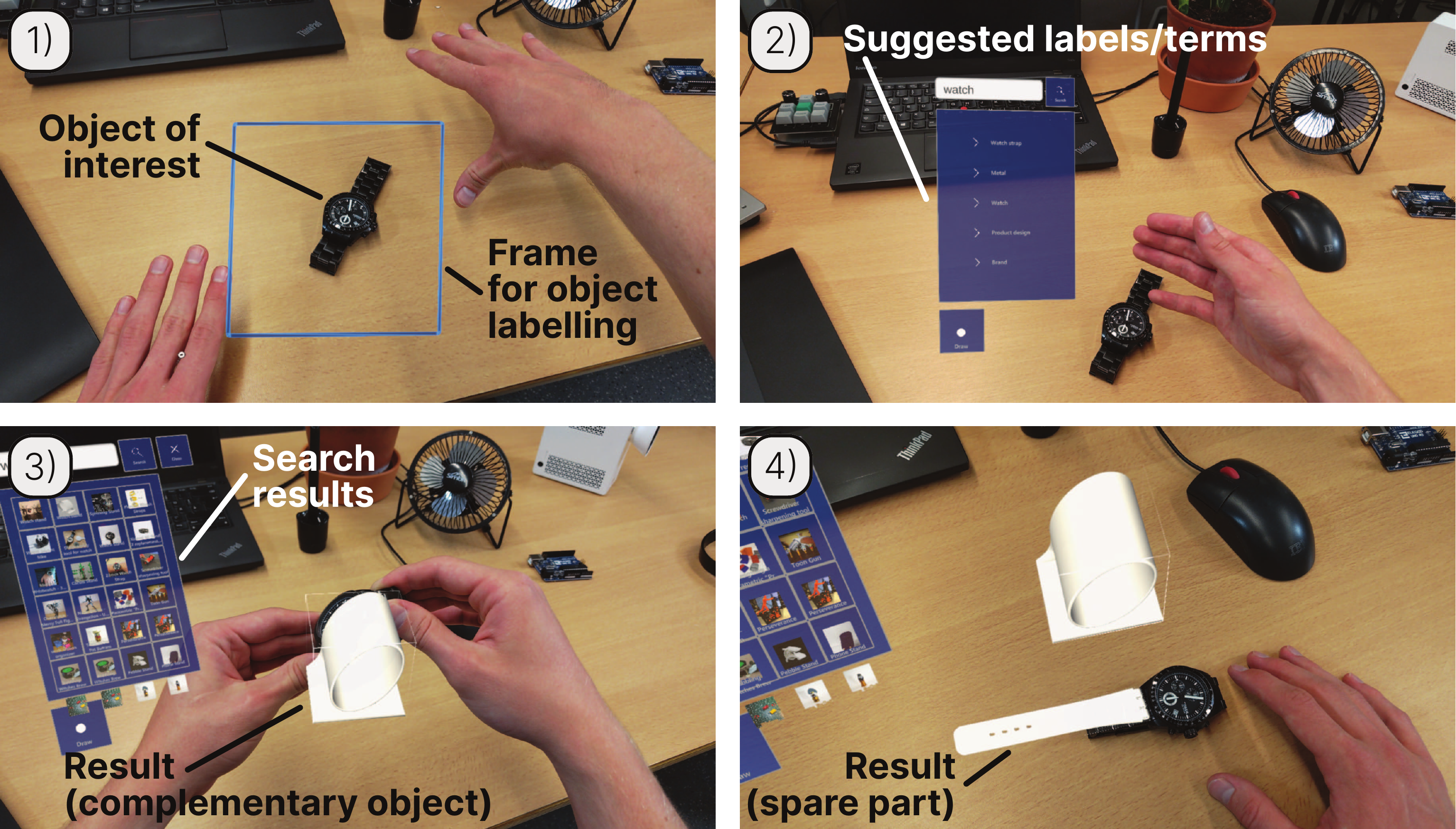}
            \caption{\textbf{Inspiration:} Using the photo-based label extraction to receive inspiration for related items. 1) Framing a wristwatch as an object of interest; 2) choosing one of the suggested terms; 3) search results related, but not identical to the framed item; 4) in-situ previewing of the artifacts.}
            \Description{ Figure 9.1:    A top-down view of a table is visible. In the center of the photo, a black wristwatch can be seen. Two hands reach into the frame and are positioned left and right of the wristwatch. The two hands are at opposite corner points of a blue square superimposed over the photo. The square frames the wristwatch.
            Figure 9.2: A floating blue user interface is augmented over the table. A text input field is filled with the word 'watch'. To the right of it, a search button is visible. Below the input field, a list of labels can be seen, containing 'watch strap', 'metal', 'watch', 'product design', 'brand'.
             Figure 9.3: The system's search UI is seen towards the left side of the photo. the term "watch" was entered as a search term and a grid of 6 by 4 results is visible (names and thumbnail images). The user holds the watch with both hands, over a white circular watch mount which is augmented over the photo.
            Figure 9.4: The watch is now laid flat on the table. A white watch strap is augmented over the real one. The circular watch mount is positioned farther on the table. A cut off part of the search UI reaches into the frame from the right.}
            \label{fig:scenario-inspiration-watch}
        \end{figure}
        
        Figure \ref{fig:scenario-inspiration-watch} depicts such a process. 
        A user frames a watch (Fig. \ref{fig:scenario-inspiration-watch}-1) and the resulting best guess is filled into the search field of \system (Fig. \ref{fig:scenario-inspiration-watch}-2).
        Additional guesses are listed below and can likewise be used for a textual query (Fig. \ref{fig:scenario-inspiration-watch}-2).
        While this is comparable to a coarse text search, it enables users to benefit from the machine's perception of the object or the scenery, ideally expanding the users' existing domain knowledge ad-hoc.
        This may include more precise terms (e.g., wristwatch instead of watch) or more general ones, in comparison to the user's wording.
        We consider this approach to be a feasible way to initiate a search with the goal of \emph{browsing} a model repository.
        The results range from objects that are meant to \emph{complement} the watch (e.g., a watch-stand -- Fig. \ref{fig:scenario-inspiration-watch}-3) or replacement parts for the watch (e.g., a watch strap -- Fig. \ref{fig:scenario-inspiration-watch}-4), all of which can be previewed in-situ.

    \subsection{Scenario 4: In-situ Features}
        With \system, users may also trace features available in the vicinity.
        This enables users to transfer aspects of their physical context to spatial queries.
        The features can be either traced coarsely and serve as a basis for a sketch extending upon them, or be traced completely to search for (geometrically) similar artifacts.
        
         \begin{figure}[b]
            \includegraphics[width=\minof{\columnwidth}{0.65\textwidth}]{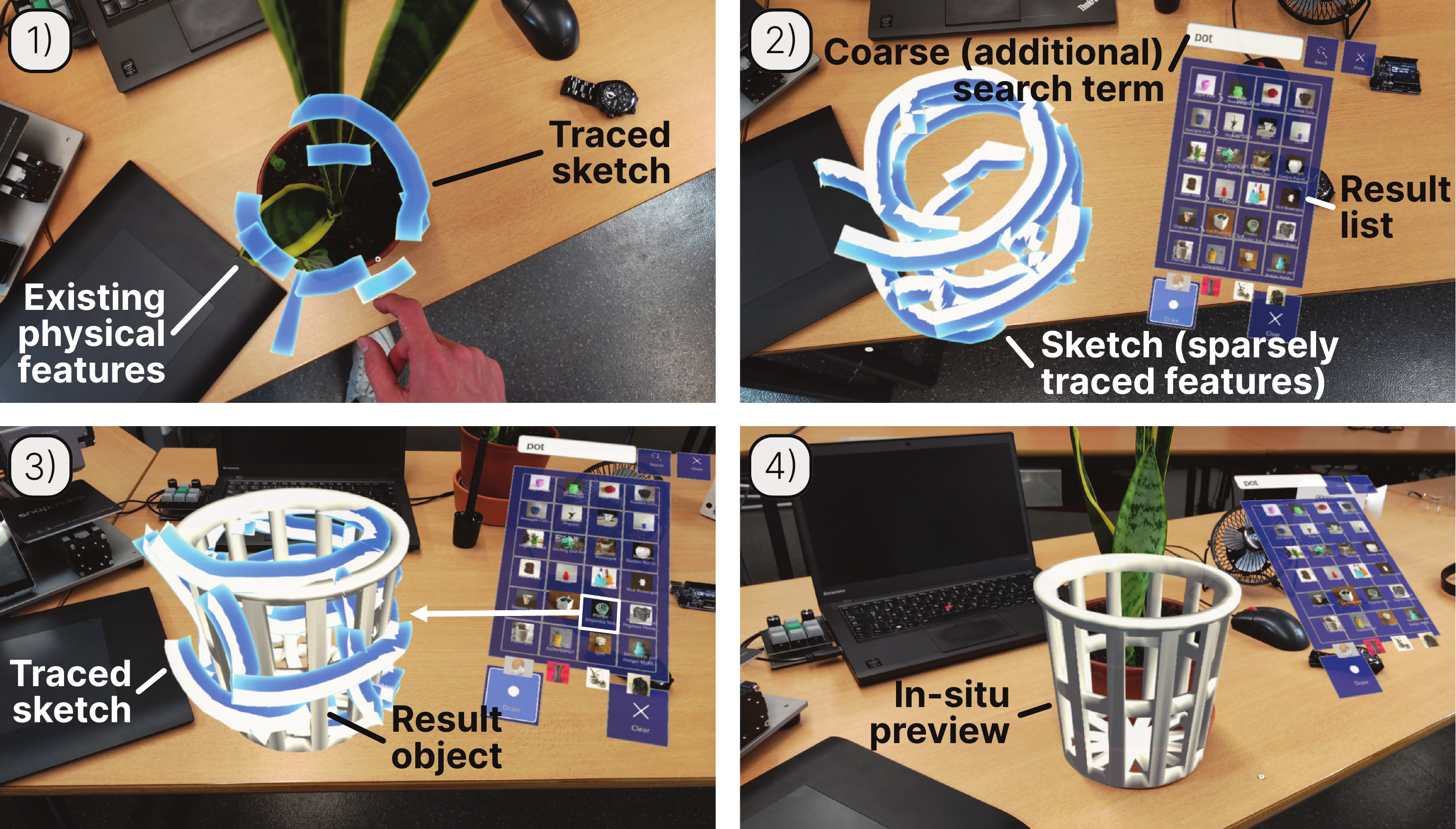}
            \caption{\textbf{Tracing} in-situ features for spatial searches. 1) coarsely tracing the coarse silhouette of the pot to be replaced; 2) removing the pot and completing the sketch; 3) submitting the query with a coarse search term, 2D result list is visible; 4) previewing in-situ.}
            \Description{Figure 10.1:    A top-down view of a table is visible. In the center of the photo, a flowerpot can be seen. A user's hand is reaching inside the frame. Around the flowerpot, blue sketch lines are seen, roughly matching the outline of the flowerpot.
             Figure 10.2:    The same table from figure 12.1 is visible. This time, the flowerpot has been removed and only the blue sketch lines remain. The system's search UI is seen towards the right side of the photo. the term "pot" was entered as a search term and a grid of 6 by 4 results is visible (names and thumbnail images)
             Figure 10.3:    A digital model of a paper bucket overlaps the blue spatial sketch on the table. The search UI is visible to the right, with one element highlighted (i.e., the one of a paper bucket). The paper bucket consists of sparse spokes and roughly matches the sketch.
            Figure 10.4:    The table scene is visible again, with the flowerpot being placed in the center. The digital model of the paper bucket is augmented over the scene again, this time overlapping the actual flowerpot. It is slightly higher than the flowerpot, but matches it in terms of radius. The search result UI is floating towards the right.}
            \label{fig:scenario-pot-trace}
        \end{figure}
        
        A user may trace an existing flowerpot he wishes to replace or improve its design (e.g., by adding a decorative cachetop or planter to it).
        The ratio of the existing pot is important for this task, so the user tries to approximate the contours of the pot as closely as possible (Fig. \ref{fig:scenario-pot-trace}-1).
        This does not necessarily leave a complete sketch of the pot.
        Afterwards, he fills some of the gaps of the outline to provide \system with volumetric features to work with (Fig. \ref{fig:scenario-pot-trace}-2). 
        Along with the search term ''planter'', the sketch search is submitted and yields decorative pots with comparable proportions to the sketch (and therefore to the original pot -- Fig. \ref{fig:scenario-pot-trace}-3).
        As the sketch was not entirely gap-free, a paper bin with sparse lines is among the first results (Fig. \ref{fig:scenario-pot-trace}-3).
        Lastly, the user may move the 3D preview of the future artifact to cover the physical pot and evaluate aesthetics or apply modifications such as scaling (Fig. \ref{fig:scenario-pot-trace}-4).
       
\section{Limitations and Discussion}
\label{sec:discussion}
    In this chapter, we want to discuss the conceptual and technical limitations of \system and the presented concept of in-situ spatial search.
    
    \subsection{Conceptual Limitations}
        Conceptually, it is important to consider the skill requirements and gaps associated with spatial sketching.
        Transferring partial requirements (e.g., tracing a physical object) to a satisfying sketch may prove hard for some users. 
        We emphasize coarse approximation of shapes, but this may not suffice for complex queries, where for instance precise measurements are required.
        This is, to a degree, compensated by other modalities available to the users (i.e., text). 
        However, this could potentially be overcome in the future with even better matching algorithms, which may consider more aspects of the environment. 
        Alternatively, this may require ways to compensate for the lack of precision through automated corrections, or novel fabrication workflows~\cite{roumenSpringFitJointsMounts2019,sunShrinkyKit3DPrinting2020}.
        
    \subsection{Technical Limitations}
        \system is, as a proof-of-concept, not without technical limitations.
        The ranking computed by the server is not impeccable, but under the candidates the system retrieves, fitting results can usually be found.
        The spatial search component is arguably slow, with one sketch-to-model comparison taking up to 2 seconds. 
        This leads to request times of approximately up to 60 seconds.
        The more artifacts are considered to be feasible candidates, the longer the search may take.
        Accelerating the server would for instance require approaches such as shape histograms~\cite{kriegelEffectiveSimilaritySearch2003}.
        To scale \system to database sizes comparable to \tv, far more optimizations and more aggressive filtering approaches are required.
        Unlike textual queries, sketch queries are far more individual and harder to cache or precompute.
        The ICP algorithm we use to align sketches and repository meshes, was originally meant to align point clouds originating from the same scene~\cite{zhangIterativePointMatching1994}.
        This is not the case with our combination, as a sketch merely approximates the target object.
        The approach of pre-filtering based on the OABB proved to be not ideal in some cases.
        If ''flat'' sketches are submitted (i.e., when the user merely traces an outline), similarly flat repository models are treated preferably.
        Automated extrusion would likely improve search performance and reduce the effort required from the users to fill volumes in their sketches.
        We argue that while there is room for improvement, coarse and ambiguous input, along with similarly coarse (but appropriate) results, may lead to more productive, creative explorations of the potential solutions to users' requirements.\\
    
    \subsection{Opportunities of In-Situ Spatial Search}
        Regardless of the aforementioned limitations, we strongly believe that \system is a proof-of-concept that demonstrates the opportunities of in-situ spatial search for future physical artifacts.
        By abstracting away from terminology, users may express their desired artifacts in a modality that is arguably more natural and fitting: in 3 dimensions and in-situ, instead of ex-situ and converted to text and labels.
        In-situ spatial search enables \textbf{Term- and Function-Abstraction}.
    	By ascribing terminology to the artifacts, they are prevented from appearing in searches not following this particular terminology (i.e., if novices to a domain are meant to express search terms).
    	In-situ spatial search may also enable \textbf{scale-invariance} in a search process.
    	With \system, we embrace the aspect that scale can be freely chosen by the user.
    	This freedom enables scale-invariant searches.
    	By normalizing all spatial input and output (while also focusing on proportions), \system sidesteps established conventions of size.
    	Future implementations may also automatically scale objects to match the user's sketch, or use the sketch as a way to filter objects by size.\\
    
    We argue that the ''ideal search'' for a physical artifact should happen in the most fitting (for the task) and easy (for the user) modality, and allow the combination of modalities to express partial requirements.
    There is no ''one size fits all'' approach to the search for physical artifacts, especially for personal use.
    A standardized artifact (e.g., an M3 screw) is easy to find, if the user knows the appropriate denomination (i.e., possesses the necessary domain knowledge).
    Likewise, \emph{precise} measurements are easier to express through text (e.g., 4cm), compared to (spatial) sketches.
    As promising as a spatial-only search system may appear, it covers and excels at a \emph{subset} of possible tasks.
    It also implicitly assumes that all users lack domain knowledge to express fitting search terms.
    If a user happens to possess the ability to express a precise search term in text, they may still achieve fitting results quickly.
    This requires systems to still offer textual means, which is what we did with \system, while also adding a transfer function from the physical context to textual queries (label extraction).
    However, proportions or geometry that are not standardized require other means to be encoded in a search, such as spatial sketches or 3D scans.
\section{Future Work}
\label{sec:futurework}
    We do not assume that 3D printing is the only domain \system may be applied to.
    Ideally, it may be a search and preview frontend to any database of physical artifacts.
    Databases that are domain-specific (e.g., furniture \cite{limParsingIKEAObjects2013}) or general artifact databases (e.g., general shopping interfaces) could be addressed by \system and the concept of in-situ spatial search.
    With the capabilities of personal fabrication devices increasing, it may apply to any physical artifact in the (distant) future.
    Additionally, it is intriguing to explore users' strategies for in-situ spatial search in a task-oriented user study.
    Technical improvements to \system are possible in terms of processing speed and the addition of further input modalities.
    With devices like the HoloLens being outfitted with a depth sensor, more detailed interaction with the environment is a promising direction.
    For instance, one may scan existing artifacts and use them as search input (3D $\rightarrow$ 3D).
    This is fairly similar to the label extraction approach (2D $\rightarrow$ 1D) present in \system, but abstracts even further from labels (\emph{Term-Abstraction}) and emphasizes geometry instead  (3D $\rightarrow$ 3D).
    This scan could also be altered with added sketch features or deformation.
    Silhouettes of artifacts are an intriguing aspect not fully considered in the \system prototype.
    Sketching outlines and applying basic operations like in revolving (for volumes with rotational symmetry), extrusion, or inflation (c.f., \cite{igarashiTeddySketchingInterface1999}) may be easier to execute than sketching the entire geometry (2D $\rightarrow$ 3D $\rightarrow$ 3D).
    
\section{Conclusion}
\label{sec:conclusion}
    We presented the concept of in-situ spatial search and \system, a mixed reality search interface to a repository of 3D-printable models.
    The prototype system embeds multiple modalities for users to search for the physical artifact they desire: sketching shapes, photo-based search (i.e., label extraction), or textual search.
    The search itself, along with the object preview, happens in-situ, at the location of the future artifact, and happens iteratively, based on input gained from the search results and the physical environment.
    We specified this concept as in-situ spatial search, and described conceptual advantages, along with walkthroughs possible with \system.
    We consider this work to be a step towards lower-effort interfaces for personal fabrication, which may emerge more as sophisticated shopping interfaces~\cite{stemasovRoadUbiquitousPersonal2021}, and less as simplified interfaces for (3D) modeling.
    
    With this work, we argue for better ways for novices and even consumers (who are not involved in personal fabrication) to search and preview future physical artifacts.
    While users with intrinsic motivation for fabrication are willing to invest time in learning and designing, the majority of people can be considered ''consumers''~\cite{stemasovRoadUbiquitousPersonal2021}.
    They are willing to benefit from unique, personal artifacts, but not willing to invest time in the process.
    For them, novel, low-effort, means for artifact \emph{retrieval}, over artifact \emph{modeling}, are required.
    The process of retrieval, in turn, should not enforce the use of a specific \emph{language} to formulate queries and demand as few transfers between (future) artifact context and the search interface as possible.

\begin{acks}
We thank \tv, \mmf, and their contributors for providing access to such vast amounts of diverse models. \newline
We furthermore thank our anonymous reviewers, who have supported us in improving the manuscript.
\end{acks}

\balance

\bibliographystyle{ACM-Reference-Format}
\bibliography{zotero}

\end{document}